\documentclass[aps,pra,twocolumn,superscriptaddress,nofootinbib,longbibliography,noshowpacs]{revtex4}
\usepackage{comment}
\usepackage{graphicx}
\usepackage{natbib}
\usepackage{amsmath}
\usepackage{mathtools}
\usepackage[breaklinks]{hyperref}
\usepackage[usenames, dvipsnames]{xcolor} 

\usepackage[normalem]{ulem}
\DeclareMathOperator*{\argminB}{argmin} 
\usepackage{lineno, blindtext}
\begin{document}
\title{Quantum-enhanced bosonic learning machine}
\author{Chi-Huan Nguyen}
\altaffiliation{These authors contributed equally to this work}
\author{Ko-Wei Tseng}
\altaffiliation{These authors contributed equally to this work}
\author{Gleb Maslennikov\footnote[2]{Present address: NKT Photonics, Bregnerødvej 144, 3460 Birkerød, Denmark}}
\author{H.~C.~J.~Gan}
\affiliation{Centre for Quantum Technologies, National University of Singapore, 3 Science Dr 2, 117543, Singapore}
\author{Dzmitry Matsukevich}
\affiliation{Centre for Quantum Technologies, National University of Singapore, 3 Science Dr 2, 117543, Singapore}
\affiliation{Department of Physics, National University of Singapore, 2 Science Dr 3, 117551, Singapore}
\date{\today}

\begin{abstract}  
Quantum processors enable computational speedups for machine learning through parallel manipulation of high-dimensional vectors~\cite{Biamonte2018}. Early demonstrations of quantum machine learning have focused on processing information with qubits~\cite{Cai2015,Travnicek2019,Havlicek2019,Johri2020a,Bartkiewicz2020,Xin2021,peters2021machine}.
In such systems, a larger computational space is provided by the collective space of multiple physical qubits.
Alternatively, we can encode and process information in the infinite dimensional Hilbert space of bosonic systems such as quantum harmonic oscillators~\cite{Braunstein2005513,Ortiz-Gutierrez2017,Fluhmann2019}. This approach offers a hardware-efficient solution with potential quantum speedups to practical machine learning problems.
Here we demonstrate a quantum-enhanced bosonic learning machine operating on quantum data with a system of trapped ions.
Core elements of the learning processor are the universal feature-embedding circuit that encodes data into the motional states of ions, and the constant-depth circuit that estimates overlap between two quantum states.
We implement the unsupervised $K$-means algorithm to recognize a pattern in a set of high-dimensional quantum states and use the discovered knowledge to classify unknown quantum states with the supervised $k$-NN algorithm.
These results provide building blocks for exploring machine learning with bosonic processors.
\end{abstract}

\maketitle
Machine learning techniques excel at finding patterns in data that are usually encoded as high dimensional feature vectors. 
As the size of these vectors grows, processing them classically requires enormous computational resources~\cite{Lloyd2013a}.
Recently proposed quantum algorithms show evidences of computational speedups, that comes from the exploitation of large Hilbert space of quantum systems~\cite{Lloyd2013a,Lloyd2014,Lloyd2014c,Rebentrost2014}.
Demonstrations of quantum machine learning algorithms implemented on discrete variables platforms have shown tremendous potential~\cite{Cai2015,Travnicek2019,Havlicek2019,Johri2020a,Bartkiewicz2020,Xin2021}. However, the number of gates and qubits must be increased for solving practical problems on near-term quantum processors.

An alternative approach is to utilize the infinite-dimensional Hilbert space of bosonic systems as the feature space for quantum machine learning~\cite{Lau2017,Schuld2019}.
One advantage of the bosonic architecture is that a single operation, such as squeezing or displacement, is formally equivalent to processing infinitely-dimensional feature vectors~\cite{Schuld2019}.
Furthermore, it allows for efficient use of physical resources. For discrete-variable schemes, approximately $\log(n)$ physical qubits are required to encode an $n$-bit data string, whereas in the alternate approach, one bosonic mode is sufficient.
As an example, in the ideal case the motion of a single trapped ion suffices to encode 3 infinitely dimensional feature vectors, whereas the internal state of an ion typically only encodes a single qubit~\cite{Ortiz-Gutierrez2017}.

\begin{figure*} [ht!]
\centering
   \includegraphics[width=\textwidth]{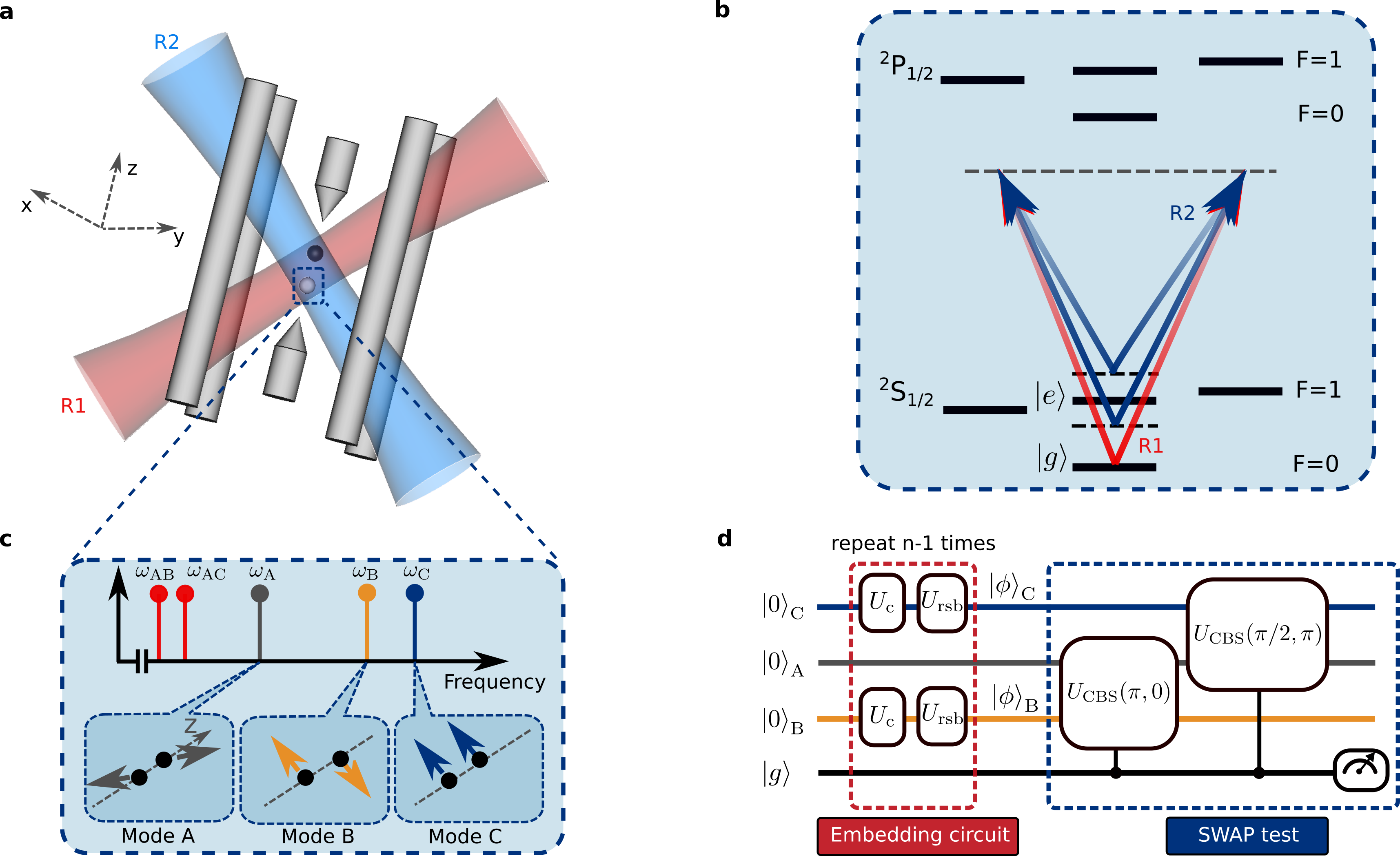}
  \caption{\label{fig:figure1} 
    \textbf{Experimental implementation}. \textbf{a}. Schematic of the linear rf trap with two trapped $^{171}$Yb$^{+}$ ions. One ion (black) is prepared in the $^{2}F_{7/2}$ metastable state such that it does not interact with optical fields applied during the experiment.\
    Two Raman beams (R1,R2) couple the motion in the radial directions with the internal states $|g\rangle,|e\rangle$.
    The $x$ and $y$ axes point along the diagonally opposite trap electrodes (Methods). \ 
    \textbf{b}. Energy level diagram of $^{171}$Yb$^{+}$. The internal states used in the experiment  are $|g\rangle = |^{2}S_{1/2},F=0,m_{F}=0\rangle$ and $|e\rangle = |^{2}S_{1/2},F=1,m_{F}=0\rangle$. The arrows represent the Raman beams.\
    \textbf{c}. Schematic of radial motional modes. $\omega_{A}$, $\omega_{B}$, and $\omega_{C}$ are frequencies of the motional modes $A$, $B$, and $C$ which denotes\
    the out-of-phase motion in the $y$ direction, out-of-phase and in-phase motion in the $x$ direction respectively.\
    $\omega_{AB}=|\omega_{A}-\omega_{B}|,\, \omega_{AC}=|\omega_{A}-\omega_{C}|$ 
    \textbf{d}. The experimental sequence to implement the bosonic learning machine. The motional modes are initialized in the ground state. The embedding circuit prepares arbitrary $n$-dimensional states in mode $B$ and $C$ by applying $n-1$ pairs of carrier ($U_\textrm{c}$) and red sideband pulses ($U_{\textrm{rsb}}$).\
    The gate sequence for the SWAP test consists of two controlled beam splitters $U_{\textrm{CBS}}(\theta,\psi)$ followed by a projective measurement of the internal state. 
}
\end{figure*}

In this letter, we report a bosonic learning machine that consists of two building blocks for encoding and processing either quantum or classical data (see Fig.~\ref{fig:figure1}d).
First, a universal embedding circuit implements a feature map $\phi$ of classical data $\boldsymbol{x}=(x_{1},x_{2}, ...,x_{n})$ to a quantum state $| \phi(\boldsymbol{x}) \rangle $. 
The map chosen in this work is amplitude encoding $| \phi_{\textrm{amp}}(\boldsymbol{x}) \rangle = 1/||\boldsymbol{x}||_2\sum\limits_{j=1}^{n} x_{j} | j-1 \rangle$, where $\{|j\rangle\}$ are the Fock states, $n$ denotes the dimension of the data, and $||\boldsymbol{x}||_2 = \sqrt{\sum\limits_{j=1}^{n} x^2_j}$ is the Euclidean norm.
The universal embedding circuit $\phi$ can be associated with a state preparation gate sequence $U_{\phi}(\boldsymbol{x})$, that prepares an arbitrary state $| \phi(\boldsymbol{x}) \rangle$ satisfying $U_{\phi}(\boldsymbol{x}) | 0 \rangle = | \phi(\boldsymbol{x}) \rangle$~\cite{Schuld2019,Lloyd2020a}.

Machine learning algorithms frequently rely on distance measures that can quantify the similarity between feature vectors~\cite{hastie2009elements}.
Here, we employ the Hilbert-Schmidt distance defined as $D_{\textrm{HS}}(|\phi (\boldsymbol{x}) \rangle,|\phi(\boldsymbol{x'}) \rangle) = \sqrt{ 2 - 2|\langle \phi(\boldsymbol{x}) | \phi(\boldsymbol{x'}) \rangle|^2}$, where $|\langle \phi(\boldsymbol{x}) | \phi(\boldsymbol{x'}) \rangle|^2$ is the overlap between two quantum states $|\phi(\boldsymbol{x}) \rangle$ and $|\phi(\boldsymbol{x'})\rangle$~\cite{Travnicek2019}.
The overlap measurement, reported throughout this work, can be efficiently realized with a constant-depth circuit implemented by a SWAP test~\cite{Filip2002,Gao2018,Jaren2020,chihuan2021}.

\begin{figure*} [ht!]
\centering
   \includegraphics[width=\textwidth]{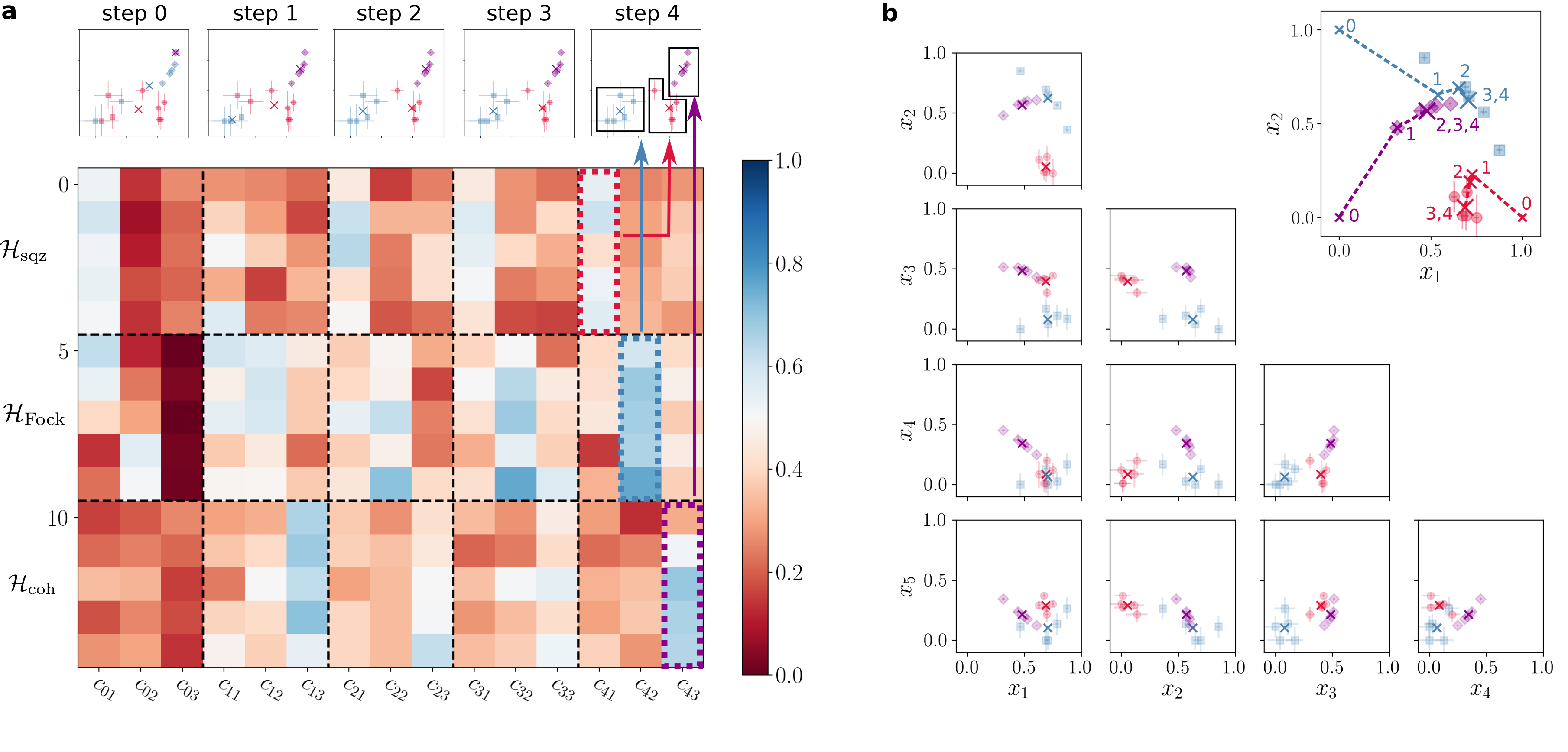}
   \caption{\label{fig:figure2}
     \textbf{ $\boldsymbol{K}$-means clustering results}. \textbf{a}. Measurement of overlap $|\langle d_{m}|{c_{lk}} \rangle|^2$ between the data states and the $k$-th centroid $|{c_{lk}} \rangle$ of the assignment step $l$.\
     The data states are the squeezed vacuum states ($\mathcal{H}_{\textrm{sqz}}$), superposition of $|0\rangle$ and $|1\rangle$ Fock states ($\mathcal{H}_{\textrm{Fock}}$), and coherent states ($\mathcal{H}_{\textrm{coh}}$).
     Top row shows the step-by-step assignment of states into clusters, visualized in the $x_3$-$x_4$ scatter plot.
      \textbf{b}. The scatter plot matrix of the centroids (crosses) and data states: $\mathcal{H}_{\textrm{sqz}}$ (circles), $\mathcal{H}_{\textrm{Fock}}$ (squares), and $\mathcal{H}_{\textrm{coh}}$ (diamonds). The classification results of $K$-means algorithm into three clusters are labeled by different colors:\
      cluster 1 (red), cluster 2 (blue), and cluster 3 (purple).
    The algorithm assigns $\mathcal{H}_{\textrm{sqz}}$ to cluster 1, $\mathcal{H}_{\textrm{Fock}}$ to cluster 2, and $\mathcal{H}_{\textrm{coh}}$ to cluster 3.
    Inset shows the process of updating the centroids visualized in the $x_1$-$x_2$ scatter plot. The error bar denotes the one $\sigma$ fit error of the state characterization (see Methods). 
 }
\end{figure*}
In this work, the bosonic learning machine comprises two trapped $^{171}$Yb$^{+}$ ions in a linear radio-frequency Paul trap, employing three motional modes denoted $A$, $B$, and $C$. We start the experiment by preparing all the modes in the ground state of motion. 
Two of the modes, $B$ and $C$, are used for storing information while $A$ acts as an auxiliary mode required for implementing the SWAP test.
The hyperfine states $|g\rangle, |e\rangle$ of one ion mediate interaction between the motional modes, while the other ion is prepared in the $^{2}F_{7/2}$ metastable state and does not interact with optical fields during the experiment (See Fig.\ref{fig:figure1}).
The Jaynes-Cumming interaction between the internal state and the motional modes enables deterministic preparation of an arbitrary state with dimension $n$ in $n-1$ steps, using methods presented in~\cite{Law1996,Ben-Kish2003}.
Alternatively, we use displacement and squeezing operations to prepare coherent and squeezed vacuum states respectively.
The SWAP test relies on controlled beam splitter gates of the form $U_{\textrm{CBS}}(\theta,\psi)=\exp[-(i / \hbar) \int_{0}^{T} \sigma_{x} H_{\textrm{BS}}(\psi,t)dt)]$ that enacts the Hamiltonian $H_{\textrm{BS}}(\psi)/\hbar=\frac{g(t)}{2}(a ^{\dagger}b e^{i\psi} +a b ^{\dagger} e^{-i\psi})$, depending on the internal state of the ions in the $\sigma_x$ basis \cite{Filip2002,Gao2018,chihuan2021}.
Here $a (a^{\dagger})$ and $b(b^{\dagger})$ are the annihilation (creation) operator of the motional modes, $g(t)$ is the coupling strength, $\psi$ an experimentally tunable phase, and $\theta=\int_{0}^{T} g(t) dt$ the effective mixing angle for a gate duration $T$ (Methods).

Given two quantum states $|\phi(\boldsymbol{x})\rangle_{B}$ and $|\phi(\boldsymbol{x'})\rangle_C$ prepared in mode $B$ and $C$ respectively, the SWAP test can be implemented with controlled beam splitter gates $U_\textrm{CBS}(\pi,0)$ acting between modes $A$ and $B$, and $U_\textrm{CBS}(\pi/2,\pi)$ between modes $A$ and $C$, followed by a projection measurement of the internal state.
The overlap between two states is proportional to the probability $P_g$ to detect the internal state in $|g\rangle$: $|_{B}\langle \phi(\boldsymbol{x})|\phi(\boldsymbol{x'})\rangle_C|^2 = |1-2P_g|$ (Fig.\ref{fig:figure1}d).
The SWAP test duration is independent from the dimension of the data and is significantly shorter than the coherence times of the motional modes~\cite{chihuan2021}. 

To highlight applications of our bosonic learning machine, we demonstrate the unsupervised $K$-means clustering and supervised $k$-NN algorithms using quantum states as the input data. 
We note that the input data may also be classical as it can be mapped to quantum states by using the universal embedding circuit.
We begin with the description of implementing the unsupervised $K$-means clustering algorithm.
Consider two parties Alice and Bob. Alice prepares and sends Bob identical copies of $M$ quantum states.
The quantum states can be expressed in the Fock basis
$| d_{m} \rangle = \sum\limits_{j=0}^{n-1} x_{m,j+1} | j \rangle$, where $m$ runs from $1$ to $M$.
In this scenario, Bob is assumed to have no \textit{a priori} information about the states that Alice prepares. Without resorting to a full tomography of the input states, Bob's task is to sort the quantum states into $K$ clusters based on similarities among the states.

The algorithm begins with Bob making a guess of the number of clusters $K$, and associates each cluster to a centroid state that are randomly initialized.
The states are assigned to the cluster that has the nearest centroid.
In step $l$ we implement a map $|d_{m}\rangle \rightarrow r^{m}_{l}=\argminB_k \{D(|d_{m}\rangle,|c_{lk} \rangle) \}$, where $|c_{lk} \rangle$ denotes the centroids corresponding to the $k-$th cluster.
Our quantum advantage comes from the reduction of complexity involved in estimating the distance of the two states $|d_{m}\rangle$ and $|c_{lk} \rangle$. This estimation takes approximately $n$ steps to perform classically, in comparison to the constant-time operation in our experiment.
The centroids are subsequently updated by Alice to be the mean of the states contained in the corresponding cluster.
The assignment and update are repeated until the clusters cease to change.

\begin{figure*} [ht!]
\centering
   \includegraphics[width=\textwidth]{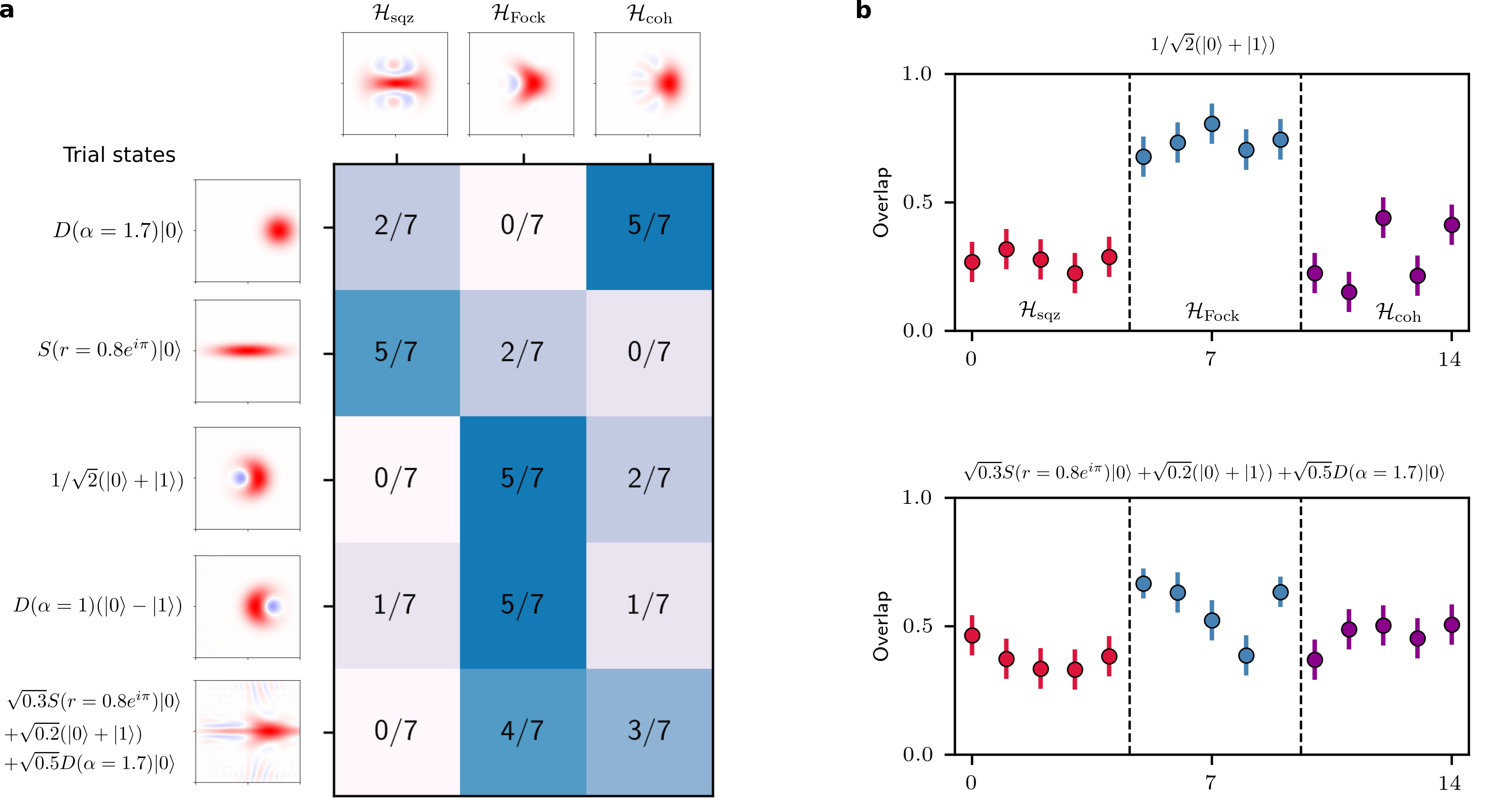}
  \caption{\label{fig:figure3} 
    \textbf{Pattern recognition of quantum states with $k$-nn algorithm}.
    The overlap between trial states and all members in each cluster were measured with SWAP test. \textbf{a}. Proportion of members in each cluster in the set of $k=7$ nearest neighbors of the trial states.
    The trial states are assigned to the cluster that has the largest proportion.\
    For visualization, the calculated Wigner functions of trial and cluster states are shown in the left column and top row, respectively.
    \textbf{b}. Measured overlap between the training data and the trial states:  $\frac{1}{\sqrt{2}}(|0\rangle + |1\rangle)$ (top) and (up to a normalization factor) $\sqrt{0.3} S(r=0.8e^{i\pi}) +\sqrt{0.2} (|0\rangle + |1\rangle) + \sqrt{0.5} D(\alpha=1.7)|0\rangle$ (bottom).\
    The error bars correspond to statistical uncertainty and represent $95 \%$ confidence intervals. Other overlap measurements can be found in the Supplementary Information.
}
\end{figure*}
Our data is a set of 15 quantum states and consists of three clusters: the set $\mathcal{H}_{\textrm{sqz}}$ of squeezed vacuum states $|r e^{i\pi} \rangle$, $\mathcal{H}_{\textrm{Fock}}$ superposition of Fock states $ \cos(\varphi /2)|0\rangle + \sin(\varphi /2)|1\rangle$, and $\mathcal{H}_{\textrm{coh}}$ coherent states $|\alpha \rangle$.
We choose the parameters as $r \in  \{1.5\,,1.1\,,1.2\,,1.3\,,1.4\} $, $\varphi/\pi \in \{0.37\,,0.45\,,0.47\,,0.72\,,0.55\}$, and $\alpha \in \{1.5\,,1.3\,,1.0\,,1.2\,,1.1\}$. Our simulations verify that $K$-means algorithm can partition this data set correctly on a classical computer.
The data states are prepared in the mode $B$.
For the first step, we choose the centroid states in mode C to be $|c_{01}\rangle = |0\rangle$, $|c_{02}\rangle = |1\rangle$, and $|c_{03}\rangle = |2\rangle$.
 After each step the centroids, truncated to dimension 5, are updated and approximately synthesized using the embedding circuit. We characterize the state generation through measurements of Rabi oscillations on the $|g\rangle|n+1\rangle$ to $|e\rangle|n \rangle$ red-sideband transition (see Methods).

Figure \ref{fig:figure2}a shows the overlap measurements between the data states and the centroids. We used on average 700 experimental runs for each measurement. With this particular choice of the initial centroids, the algorithm attains a stationary assignment after four steps.
The final classification result is given in Fig.~\ref{fig:figure2}b, where the scatter plot of the amplitudes $x_{j}$ of the data states are shown for $j$ up to 5.
The algorithm associates cluster 1 to $H_{\textrm{sqz}}$, cluster 2 to  $H_{\textrm{Fock}}$, and cluster 3 to $H_{\textrm{coh}}$.
The SWAP test measurements, albeit imperfect, produce classification results that agrees perfectly with the prior knowledge about the states sent to Bob.
Stationary assignment of the centroids is evident from the step-by-step evolution shown in the inset of Fig.~\ref{fig:figure2}b.
The assignment results after each step can be found in Extended Data.

We next demonstrate the $k$-NN algorithm to classify unknown trial states, by using the three labelled clusters from the $K$-means results as training states.
The traditional approach to solving the problem of classifying unknown states is to perform a complete tomography, which requires a number of measurements that grows exponentially with the state dimension $n$~\cite{Haah2017}.
However, if complete information regarding the state is unnecessary, $k$-NN algorithm could offer an efficient alternative, especially when $n$ is large. With a constant depth SWAP test, computational complexity of the $k$-NN becomes independent of $n$. 

The $k$-NN algorithm requires estimation of the overlap between each trial state and all training states.
We identify the $k$ states having highest overlap with the trial state. 
The trial state is then assigned to the cluster that labels the majority of $k$ states.
Figure~\ref{fig:figure3} shows the experimental results of $k$-NN classification with $k=7$ for a variety of trial states.
The first three trial states utilize the same operation that prepares the three clusters, respectively, and are observed to be correctly classified.
We further carry out the classification of trial states that are prepared with other operations (See Methods).
Such examples are the two states $D(\alpha=1)(|0\rangle -|1\rangle)$ and $\sqrt{0.3}S(r=0.8e^{i\pi})|0\rangle + \sqrt{0.2}(|0\rangle +|1\rangle)+\sqrt{0.5}D(\alpha=1.7)|0\rangle$ (up to a normalization factor), that are both classified to cluster 2. 
These results indicate that the outcome of the algorithm is determined by overlap of the quantum states, which can be visualized by the similarity of their Wigner functions.
This can be interpreted as a form of Wigner function pattern recognition, much like applying $k$-NN to handwriting recognition in classical ML~\cite{hastie2009elements}.

In conclusion, we realize a bosonic learning machine using motional modes in a system of trapped ions. 
The core subroutine of the machine is the efficient SWAP test that estimates the overlap of two motional modes with a constant gate duration, a feature that is preferred in solving problems with large dimension such as quantum support vector machines. 
We expect that the experimentally accessible dimensions can be further increased by improving various technical aspects of the setup, such as the motional coherence time and coupling strength between modes.
The same bosonic learning machine can also be used to implement several quantum kernels proposed in~\cite{Schuld2019} and on different experimental platforms such as superconducting microwave cavities~\cite{Hofheinz2009,Gao2018}.
These results provide a key primitive for developing advanced quantum-enhanced machine learning algorithms in infinite-dimensional feature spaces.
\begin{acknowledgments}
We thank P.T. Le, H.N. Le, and L.C. Kwek for discussions.
This research is supported by the National Research Foundation, Prime Ministers Office, Singapore, and the Ministry of Education, Singapore, under the Research Centers of Excellence program and NRF Quantum Engineering Program (Award QEP-P4)
\end{acknowledgments}
\begin{center}
\textbf{Author contributions}
\end{center}

C.H.N. and D.M. conceived the experiment. C.H.N. K.W.T, G.M. and H.C.J.G. built and maintained the experimental apparatus.
C.H.N. and K.W.T performed the experiments and analyzed the experimental data with assistance from G.M. and H.C.J.G.
All authors contributed to the manuscript preparation.

\begin{center}
\textbf{Data availability}
\end{center}

The data that support the findings of this study are available from the corresponding authors on reasonable request.

\begin{center}
\textbf{Author Information}
\end{center}

The authors declare no competing financial interests. Correspondence and requests for materials should be addressed to C.H.N. (ngchihuan@gmail.com) or D.M. (phymd@nus.edu.sg).

\begin{center}
    \textbf{Methods}
\end{center}

\textbf{Experimental setup.}
We trap two $^{171}$Yb$^{+}$ ions in a four-rod linear radio-frequency Paul trap with single-ion secular trap frequencies $(\omega_x,\omega_y,\omega_z) = 2 \pi \times (1.274, 0.945, 0.519)$\,MHz.
The normal motional modes $A$, $B$, and $C$ are the out-of-phase motion in the $y$ direction, out-of-phase motion and in-phase in the $x$ direction with frequencies $(\omega_{A},\omega_{B},\omega_{C})=2\pi \times (0.782, 1.159, 1.274)$\,MHz, respectively.
The transverse trapping frequencies are actively stabilized and drift less than 100 Hz/hour.

One of the ions, pumped to the metastable state $^{2}F_{7/2}$ by driving the $^{2}D_{3/2} \rightarrow$ $3[1/2]^{\circ}_{3/2}$ transition at $398.98$\,nm, does not interact with optical fields and is referred to as the ``dark'' ion.
At the beginning of all experiments, the other ion is Doppler cooled by a 369.53\, nm laser red-detuned from the $|^{2}S_{1/2},F=1\rangle$ to $|^{2}P_{1/2},F=0\rangle$ transition, and sympathetically cools the ``dark'' ion.
The two hyperfine states used in the experiment are $|g\rangle = |^{2}S_{1/2},F=0,m_{F}=0\rangle$ and $|e\rangle = |^{2}S_{1/2},F=1,m_{F}=0\rangle$, which are separated by $\omega_0 /2\pi=12.643$\,GHz.
The ion is initialized via optical pumping in the state $|g\rangle$.
Resonance fluorescence state detection technique is used to detect the ion's internal state~\cite{Olmschenk2007}.

A frequency-doubled, mode-locked Ti:Sapphire laser with a central wavelength of 372.85\,nm is used to drive the stimulated Raman transitions. 
The mode-locked laser has a pulse duration of 3\,ps and a repetition rate of 76.2\,MHz.
Two beams, R1 and R2 (Fig.~\ref{fig:figure1}a,b) with an average power of 60mW each, are sent through acoustic-optical modulators (AOMs) and focused to a beam waist of $\sim15\,\mu$m at the ions.
The beams R1 and R2 form 45$^{\circ}$ and 135$^{\circ}$ angles with respect to the $z$-axis (Fig.~\ref{fig:figure1}a).
The detuning of the Raman beams are controlled by the AOMs.

We begin all experiments with all the radial motional modes prepared in the ground state (residual $\bar{n} \leq 0.05$) using 10\,ms Sisyphus cooling followed by 250 cycles of Raman sideband cooling.

\textbf{Controlled beam splitter gates.}
We implement controlled beam splitter gates $U_\textrm{CBS}$ between modes $A$ and $B$ by applying a bichromatic Raman beatnote at $ \omega_0 \pm \omega_{AB}$, where $\omega_{AB} = |\omega_{A} - \omega_{B}|$.
The resulting Hamiltonian is
\begin{equation}
H_{I} = \frac{\hbar g(t)}{2} \sigma_{x}( a^\dagger b e^{-i\psi} + a b^\dagger e^{i\psi}) =  \sigma_x H_{\textrm{BS}},
\end{equation}
where $\psi = \frac{\psi_b - \psi_r}{2}$, and $\psi_r$ and $\psi_b$ are the phases of the bichromatic fields.
We let $\sigma_{x}=(e^{-i\psi_S}\sigma_+ + e^{i\psi_S}\sigma_-)$, where $\psi_{\textrm{S}} = -\frac{\psi_r + \psi_b}{2}$, and $\sigma_+ = |e\rangle \langle g |$, $\sigma_- = |g\rangle \langle e |$.
$g(t)$ is the coupling strength.

The controlled beam splitter gate corresponds to the unitary evolution $U_{\textrm{CBS}}(\theta,\psi)=\exp[-(i / \hbar) \int_{0}^{T} \sigma_{x} H_{\textrm{BS}}(\psi,t)dt)]$, where $\theta=\int_{0}^{T} g(t) dt$.
In the experiment, we obtain a coupling strength $g/2 \pi \sim 680$\,Hz.
This corresponds to implementing a beam splitter operation $U_{\textrm{CBS}}(\theta = \pi/2)$ in $T \sim 365$\,$\mu$s, much shorter than
the motional phase coherence time for the superposition of $|0\rangle+|1\rangle$ which is approximately $10$\,ms. 

\textbf{Preparation and characterization of quantum states.} 
The universal embedding circuit relies on a series $n-1$ pairs of carrier and red-sideband:
\begin{equation}
\prod_{l=1}^{n-1} U_{\textrm{rsb}}(\tau_{l},\Upsilon_{l})U_{c}(t_{l},\upsilon_{l} + \upsilon^{'}_{l})
\end{equation}
where $t_{l}$ ($\tau_{l}$) and $\upsilon_{l}$ ($\Upsilon_{l}$) are the duration and phase for the $l$-th carrier (red-sideband) pulse that are determined by the Eberly-Law algorithm~\cite{Law1996}. 
The additional phases $\upsilon^{'}_{l}$ compensate for the rotation of the internal states due to the AC Stark shifts of the red-sideband pulses (See Supplementary). 
The embedding circuit is used to prepare the centroid states of $K$-means, the cluster state $\mathcal{H}_{\textrm{Fock}}$ of $K$-means experiment, and the trial state (up to a normalization factor) $\sqrt{0.3}S(r=0.8e^{i\pi})|0\rangle + \sqrt{0.2}(|0\rangle +|1\rangle)+\sqrt{0.5}D(\alpha=1.7)|0\rangle$ of $k$-NN experiment.
We use spin-dependent force to prepare all coherent states and the squeezed vacuum states~\cite{Leibfried2003}. 
For the trial state $D(\alpha=1)(|0\rangle - |1\rangle)$, we first prepare $|0\rangle -|1\rangle$, followed by a displacement of $\alpha=1$.

The real and positive coefficients $x_{j+1}$ of the data states $|d_m\rangle$, shown in Fig.~\ref{fig:figure2}b, are the square root of population distribution $p(j)$, where $p(j) = |\langle d_m|j\rangle|^2$.
The population distribution is determined from a Fourier transform of the time evolution while driving the red-sideband transition
\begin{equation}
\label{eq:Pe}
P_e(\tau)=\sum\limits_{j=1}^{\infty}p(j)e^{-\sqrt{j}\gamma\tau}(1-\cos{(\sqrt{j}\Omega_\textrm{rsb} \tau}))/2
\end{equation}
where $P_e$ is the probability to detect the ion in the state $|e\rangle$, $\Omega_\textrm{rsb}$ is the Rabi frequency of the red-sideband, and $\gamma$ is the decoherence rate of motional states.
To determine $p(j)$ a fit constrained to be non-negative and $\sum_{j=0}^{6}p(j)=1$ is applied to Eq.~\ref{eq:Pe} for $j=0$ to $6$.

\bibliographystyle{apsrev4-1}

\bibliography{qml.bib}
\clearpage
\widetext
\begin{center}\textbf{\large Supplementary Material: Experimental quantum-enhanced machine learning over infinite dimensions}\end{center}
\setcounter{equation}{0}
\setcounter{figure}{0}
\setcounter{table}{0}
\setcounter{page}{1}
\makeatletter
\renewcommand{\theequation}{S\arabic{equation}}
\renewcommand{\thefigure}{S\arabic{figure}}
\renewcommand{\bibnumfmt}[1]{[S#1]}
\renewcommand{\citenumfont}[1]{S#1}

\section{Delay time between state preparation and the SWAP test}
The two steps required for operation of the bosonic learning machine are the state preparation and the overlap measurement using the SWAP test. However, the quantum state of a harmonic oscillator evolves in time as $|\Psi\rangle = \sum_j c_j e^{-i \omega j t}|j\rangle $ where $\omega$ is the motional frequency. If the frequencies of motional modes $\omega_B$ and $\omega_C$ are different, care must be taken to calibrate waiting time between preparation and measurement to ensure that intended states are passed to the SWAP test. 
We set the delay time as $nT + \textrm{offset}$. Otherwise, for example, the overlap of the two coherent states in modes $B$ and $C$ will oscillate as a function of the delay between the two steps, with a period of $T=2 \pi / |\omega_C - \omega_B|$.

The trap frequency measurements done by scanning the motional sidebands are affected by the Stark shifts of the internal states. To obtain a frequency difference between the modes which is free from AC Stark shifts we measure the overlap $|\langle \Psi|\Phi_{1}\rangle|^2$ as a function of the delay time $\tau$, where $| \Psi \rangle = |\Phi_{1}\rangle = 1/\sqrt{4}(|0\rangle + |1\rangle +|2\rangle +|3\rangle) $.
Figure S1 shows the results of overlap measurement. 
The period $T$ is obtained from the time difference between two peaks in the $|\langle \Psi|\Phi_{1}\rangle|^2$ overlap measurement. 
We estimate $T \sim 8.2\, \mu$s.
\begin{figure*}[!h]
\centering
   \includegraphics{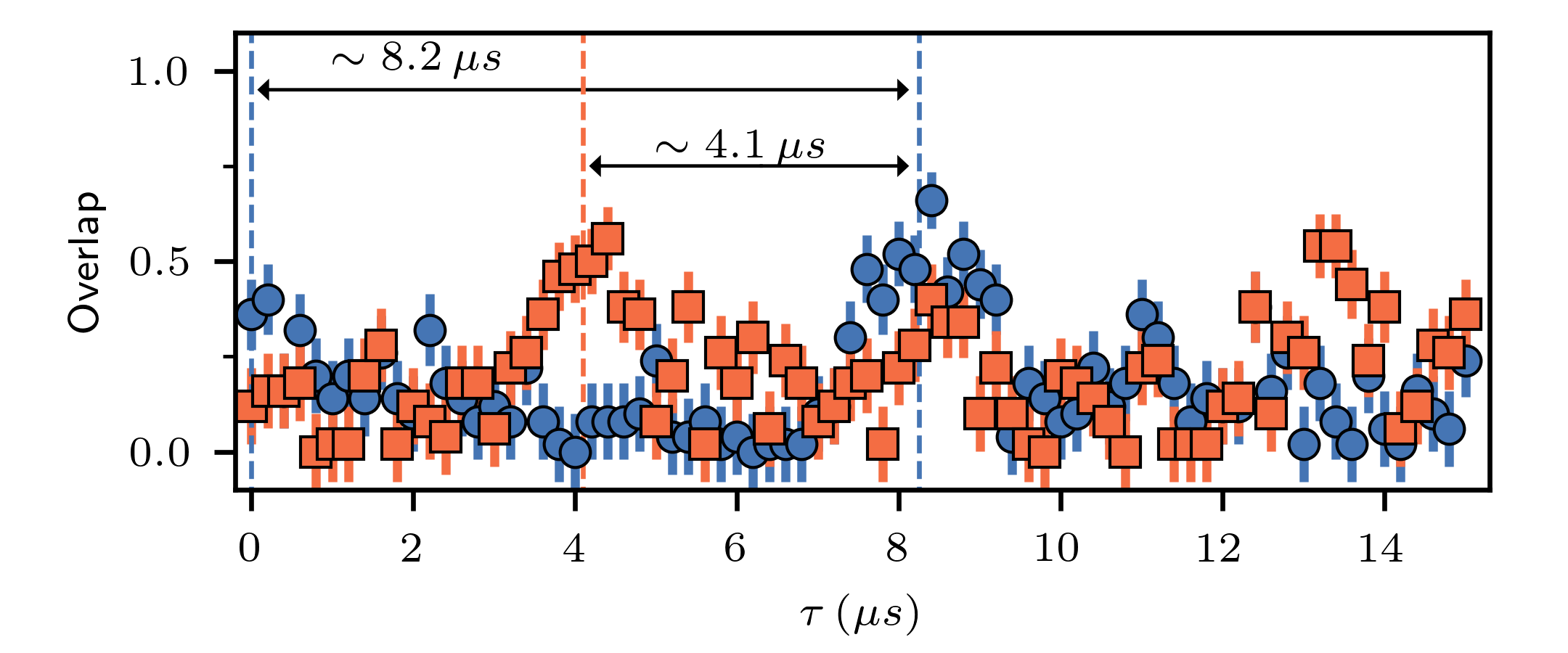}
  \caption{\label{fig:figureS1} 
Measurement of $|\langle \Psi|\Phi_{1}\rangle|^2$ (blue circles) and $|\langle \Psi|\Phi_{2}\rangle|^2$ (orange squares) as a function of waiting time $\tau$, where $| \Psi \rangle = |\Phi_{1}\rangle = 1/\sqrt{4}(|0\rangle + |1\rangle +|2\rangle +|3\rangle) $ and $|\Phi_{2}\rangle = 1/\sqrt{4}(|0\rangle - |1\rangle +|2\rangle -|3\rangle)$.
}
\end{figure*}

\section{Additional details on state preparation}
The Fourier analysis of the time evolution while driving the red sideband can measure only the diagonal terms (phonon distribution) of the density matrix of the states.
To verify that the universal embedding circuit can prepare states with desired off-diagonal terms, we carry out the following experiment. We prepare $|\Phi_{2}\rangle = 1/\sqrt{4}(|0\rangle - |1\rangle +|2\rangle -|3\rangle)$ and compare the two overlaps $|\langle \Psi|\Phi_{1}\rangle|^2$ and $|\langle \Psi|\Phi_{2}\rangle|^2$ as a function of waiting time $\tau$. Simulations predict that the second overlap $|\langle \Psi|\Phi_{2}\rangle|^2$ is shifted in time, relative to $|\langle \Psi|\Phi_{1}\rangle|^2$, by $T/2$ if the states with correct off-diagonal terms are prepared. The experimental results shown in Fig. S1 agree with calculations.
\begin{figure*}[!h]
\centering
   \includegraphics{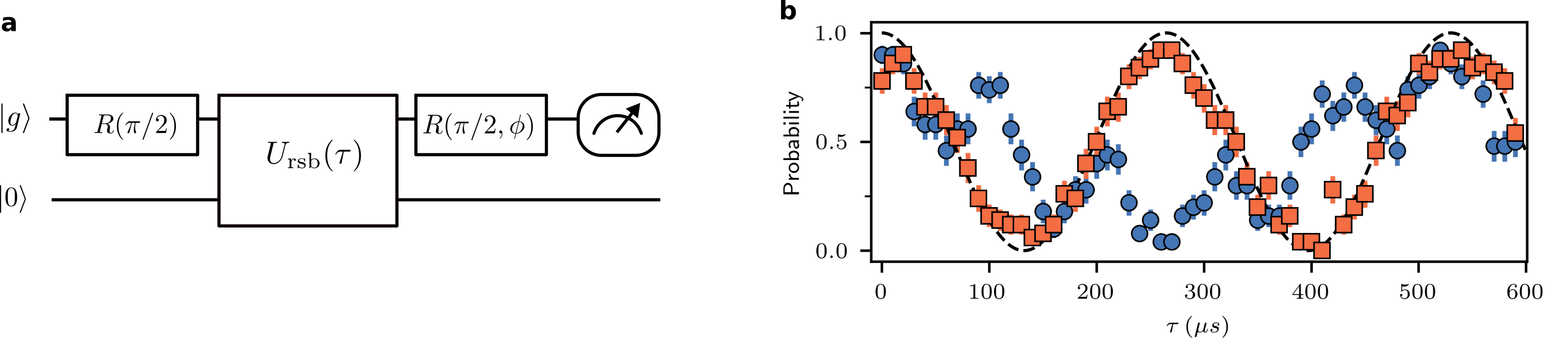}
  \caption{\label{fig:figureS1} 
  \textbf{a.} Circuit diagram for measuring the AC Stark shift induced by the inserted rsb pulse $U_\textrm{rsb}$. $R(\pi/2,\phi)$: $\pi/2$ rotation pulses with phase $\phi$  of the internal states.
  \textbf{b.} Probability of detecting the internal state in $|e\rangle$ as a function of the red-sideband pulse duration $\tau$ with $\phi=0$ (blue circles) and $\phi=- 2\pi \times 5.9\tau$\,kHz (orange squares). The dashed line is the calculated probability when the AC Stark is compensated.
}
\end{figure*}

The red-sideband pulse $U_{rsb}(\tau_{l},\Upsilon_{l})$ induces an AC Stark shift $\hbar \Delta \omega_{\textrm{AC}}$ that gives rise to a rotation $U=e^{-i \Delta \omega_{\textrm{AC}} \tau_l \sigma_z}$ of the internal states. 
To compensate for this rotation, an additional phase $\upsilon^{'}_{l+1} = \sum_{l' \leq l} \Delta \omega_{\textrm{AC}} \tau_{l'}$ is included to the next carrier pulses $U_{c}(t_{l},\upsilon_{l} + \upsilon^{'}_{l})$. We determine $\Delta \omega_{\textrm{AC}}$ using the Ramsey experiment with an inserted red-sideband pulse described in Figure S2a. The second $\pi/2$ rotation pulse has a phase $\phi$. We vary the red-sideband pulse duration $\tau$ while monitoring the probability of detecting the internal state in $|e \rangle$. When $\phi$ is set to $-\Delta \omega_{\textrm{AC}} \tau$ and $\Delta \omega_{\textrm{AC}} \sim 2\pi \times 5.9\,$KHz, the measurement result (Fig. S2b) agrees with the expectation when the AC Stark shift is compensated.

\section{Extended Data}

\begin{figure*}[!h]
\centering
  \includegraphics{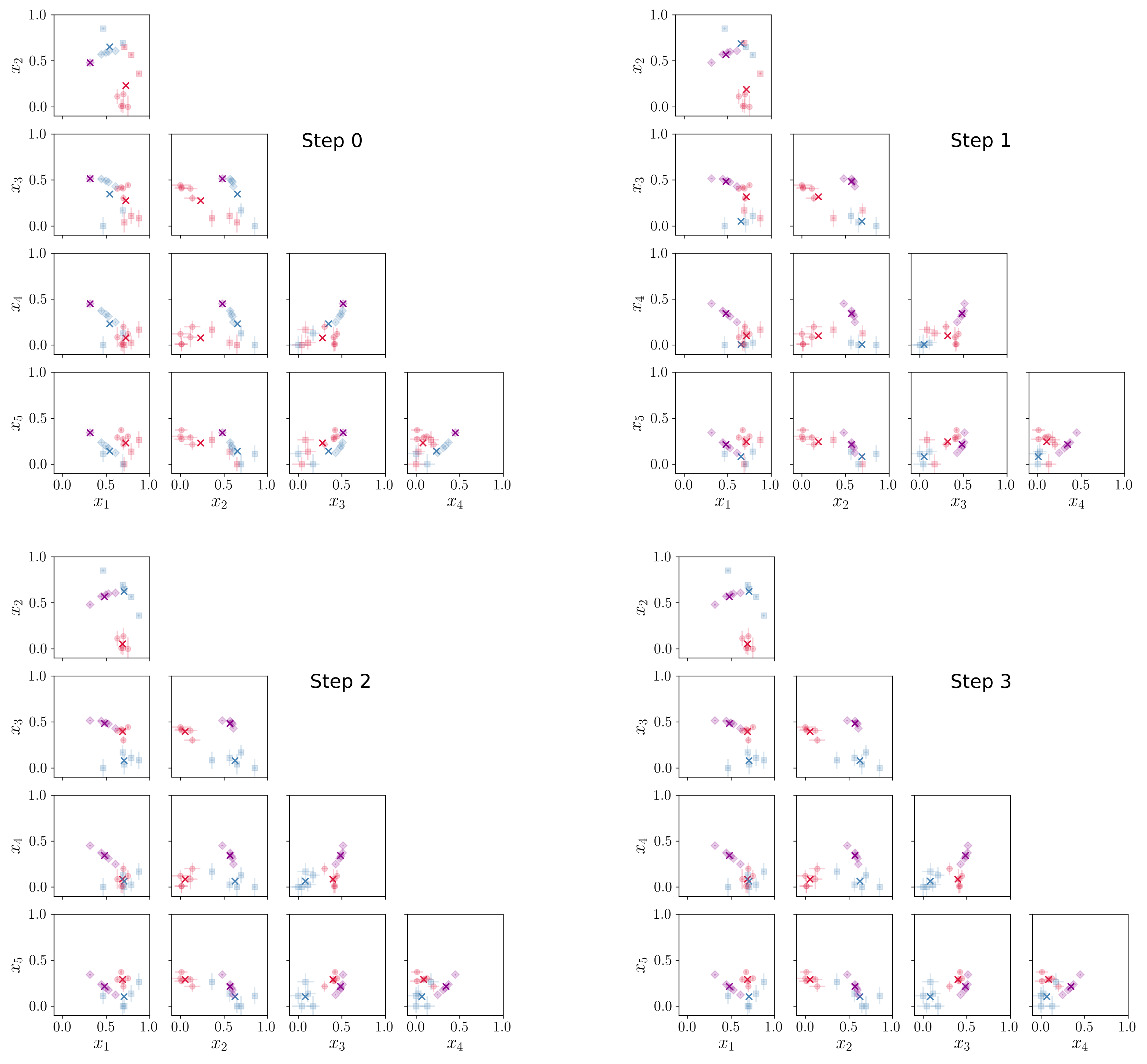}
  \caption{\label{fig:figureS3} 
  Step-by-step cluster assignment of $K$-mean algorithm visualized in scatter plot matrix of the centroids (crosses) and data states: $\mathcal{H}_{\textrm{sqz}}$ (circles), $\mathcal{H}_{\textrm{Fock}}$ (squares), and $\mathcal{H}_{\textrm{coh}}$ (diamonds). The error bar denotes the one $\sigma$ fit error of the state characterization.
}
\end{figure*} 
\newpage
\begin{figure*}[!h]
\centering
  \includegraphics{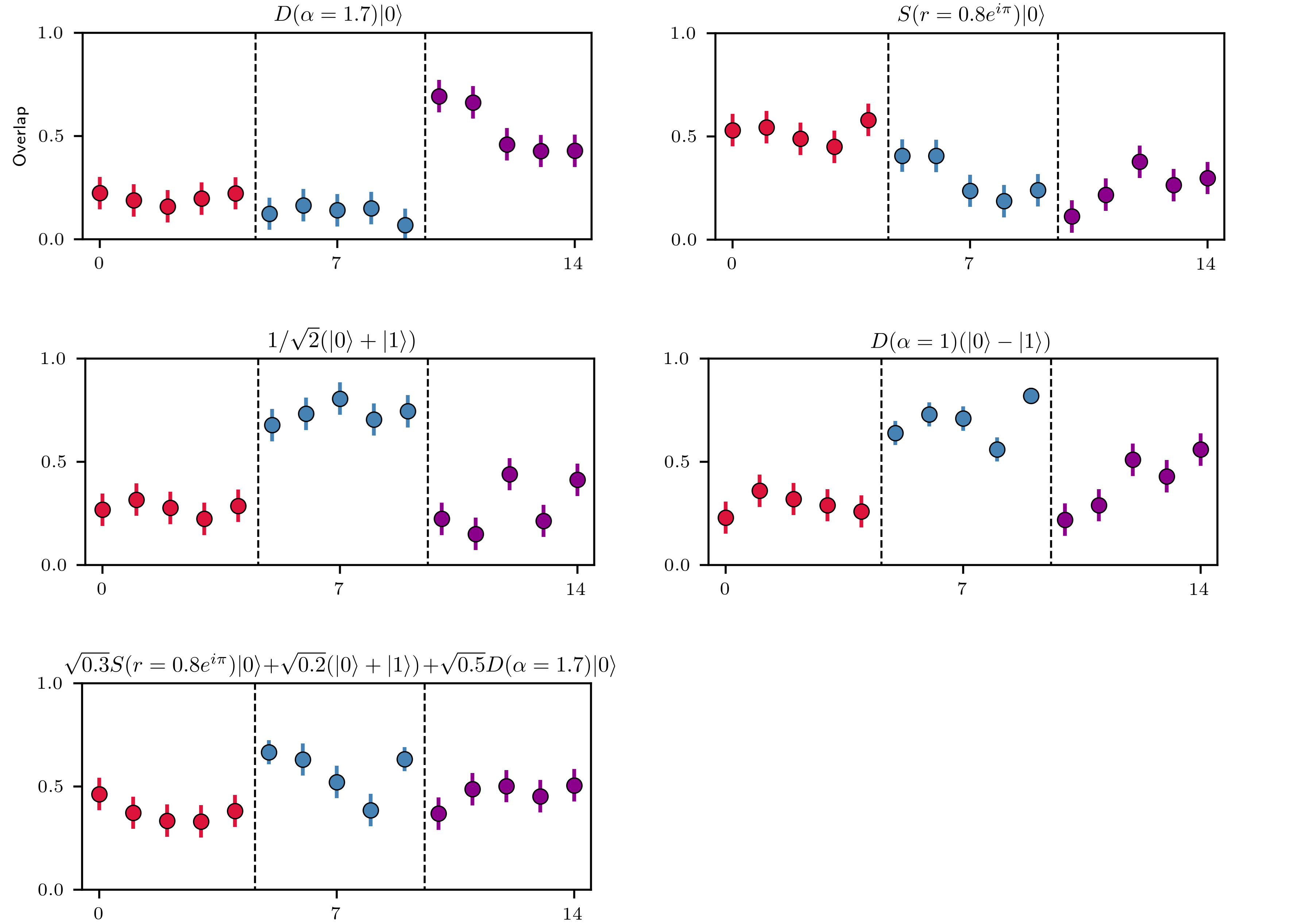}
  \caption{\label{fig:figureS4} 
 The overlap between trial states and all members in each cluster in the $k$-nn experiment. The error bars correspond to statistical uncertainty and represent $95 \%$ confidence intervals.
}
\end{figure*} 

\newpage
\begin{figure*}[!h]
\centering
  \includegraphics{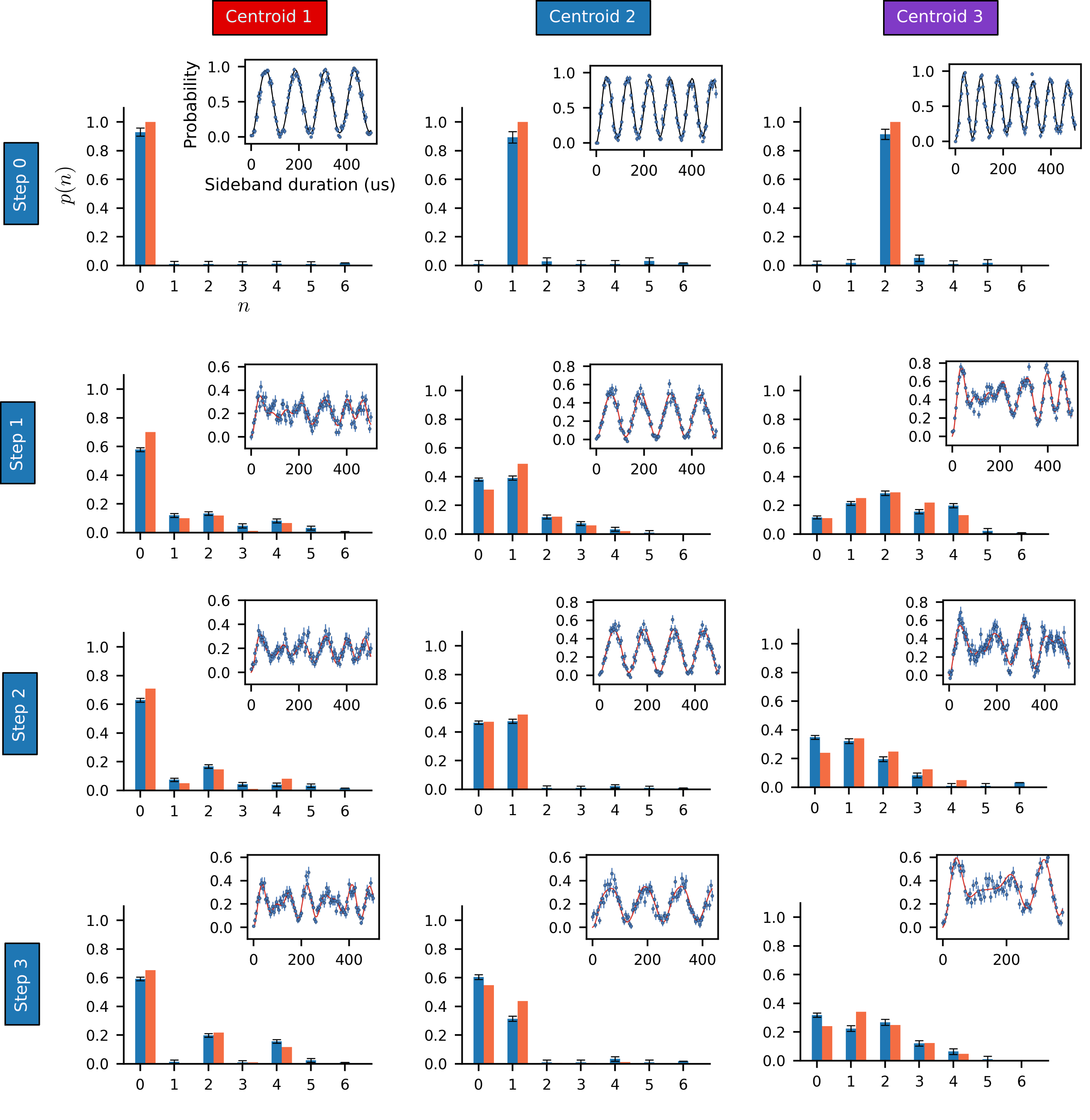}
  \caption{\label{fig:figureS5} 
 Population distribution of the centroid states in the Fock basis. Blue sideband pulses are used to analyze centroid states of step 0 and red sideband pulses to analyze the rest of the centroid states. Insets show the probability to detect the internal state in $|e\rangle$ as a function of the pulse duration. The red curves are the fits to the population using Eq.~\ref{eq:Pe}. The orange bars show the desired population distribution.
}
\end{figure*} 
\end{document}